\documentclass[12pt,preprint]{aastex}

\begin{document}

\shorttitle{Tidally Crushed White Dwarfs}

\shortauthors{Rosswog, Ramirez-Ruiz \& Hix}

\title{Atypical Thermonuclear Supernovae from Tidally Crushed
  White Dwarfs}

\author{Stephan Rosswog\altaffilmark{1}, Enrico
  Ramirez-Ruiz\altaffilmark{2}, and William R. Hix\altaffilmark{3}}
\altaffiltext{1}{School of Engineering and Science, Jacobs University
  Bremen, Campus Ring 1, 28759 Bremen, Germany; s.rosswog@jacobs-university.de}
\altaffiltext{2}{Department of Astronomy and Astrophysics, University
  of California, Santa Cruz, CA 95064; enrico@ucolick.org}
\altaffiltext{3}{Physics Division, Oak Ridge National Laboratory, Oak Ridge,
  TN37831-6374; raph@phy.ornl.gov}

\begin{abstract} 
Suggestive evidence has accumulated that intermediate mass black holes
(IMBHs) exist in some globular clusters. As stars diffuse in the
cluster, some will inevitable wander sufficiently close to the hole
that they suffer tidal disruption. An attractive feature of the IMBH
hypothesis is its potential to disrupt not only solar-type stars but
also compact white dwarf stars. Attention is given to the fate of
white dwarfs that approach the hole close enough to be disrupted and
compressed to such extent that explosive nuclear burning may be
triggered. Precise modeling of the dynamics of the encounter coupled
with a nuclear network allow for a realistic determination of the
explosive energy release, and it is argued that ignition is a natural
outcome for white dwarfs of all varieties passing well within the
tidal radius. Although event rates are estimated to be significantly
less than the rate of normal Type Ia supernovae, such encounters may
be frequent enough in globular clusters harboring an IMBH to warrant a
search for this new class of supernova.
\end{abstract}
 
\keywords{black hole physics -- supernova: general -- white dwarfs --
  globular clusters: general}

\section{Introduction}
The conjecture that there may be an intermediate mass black hole
(IMBH) in some globular clusters has gained credence over the past few
years. There is some dynamical evidence for a mass concentration
within the central regions of some globular clusters. In particular,
the dynamics of stars in the inner regions of nearby clusters such as
M15 and G1 suggest the presence of black holes with masses of the
order of $10^{3}$ and $10^{4}\;M_\odot$, respectively
\citep{ge2002,ge2003,geb2002,geb2005}. This evidence remains rather
controversial, partly because the velocity dispersion profiles can be
reproduced without invoking the presence of an IMBH.  Recently,
somewhat more ambiguous evidence has arisen for the presence of IMBHs
in young star clusters, where ultraluminous, compact X-ray sources
(ULXs) have been preferentially found to occur
\citep{ze2002,po2006}. Their high luminosities suggest that they are
IMBHs rather than binaries containing a normal stellar mass black
hole. Studies based on detailed $N$-body simulations
\citep{po2004,patruno06} of the evolution of a young star cluster in
M82, the position of which coincides with an ULX, give further
credence to this idea. While none of these arguments in itself carries
enough weight to be convincing, the fact that they are so different in
character does suggest that the existence of IMBHs in some young star
clusters should be taken seriously.

One would like some independent corroboration of the IMBH hypothesis,
or, conversely, some way of ruling it out.  Stellar disruption
potentially offers such a test. As stellar orbits diffuse in phase
space, it therefore seems inevitable that some will wander
sufficiently close to the hole that they suffer tidal disruption. A
star interacting with a massive black hole cannot be treated as a
point mass if it gets so close to the hole that it becomes vulnerable
to tidal deformations \citep{fr1976}. The tidal radius, $r_{\rm T}$,
defined as the distance within which a stars gets tidally disrupted,
obviously has a value that depends on the type of star being
considered. For a solar-type star it is 
\begin{equation}
r_{\rm T} = \left(\frac{M_{\rm h}}{M_{\odot}} \right)^{1/3} r_{\odot}
\simeq 5 \times 10^{12} M_{\rm h,6}^{1/3} r_{\odot} M_{\odot}^{-1/3}\;{\rm
  cm},  
\end{equation}
while for a degenerate white dwarf is roughly 
\begin{equation}
r_{\rm T} \simeq 10^{11} M_{\rm h,6}^{1/3} r_{\rm wd,9} M_{\rm
  wd,0.6}^{-1/3} \; {\rm cm}.  
\end{equation}
Here $r_{\rm wd,9}$ and $M_{\rm wd,0.6}$ denote the white dwarf radius
and mass in units of $10^{9}$ cm and $0.6 M_\odot$, respectively. Note
that for white dwarfs, the tidal radius is indeed inside the
gravitational radius, $r_{\rm g} \simeq 1.5\times 10^{11} M_{\rm
  h,6}\;{\rm cm}$, for black holes masses exceeding
\begin{equation}
M_{\rm h, lim} \sim 10^{5} r_{\rm wd,9}^{3/2}M_{\rm wd,0.6}^{-1/2}\;
M_\odot. 
\end{equation}
For this reason, tidal disruption of white dwarfs offers a unique
diagnostic for the presence of an IMBH
\citep{lu1989,wi2004,ba2004}. This paper explores the observational
manifestations of such phenomena, with particular reference to white
dwarfs that approach the hole close enough to be disrupted and
severely compressed.

\section{Numerical Methods}
The observational consequences of stellar disruption depend on what
happens to the debris \citep{re1988,ev1989}. To quantify this, we have
performed detailed three-dimensional, hydrodynamical calculations. The
gas dynamics is coupled with a nuclear network to explore the effects
of explosive nucleosynthesis during the strong compression phase.

{\it Hydrodynamics}. The smoothed particle hydrodynamics method (SPH)
is used here to solve the equations of hydrodynamics. Due to its
Lagrangian nature SPH is perfectly suited to follow tidal disruption
processes during which the corresponding geometry, densities and time
scales are changing violently. The SPH-formulation used in this study
follows closely the one described in \citet{be1990}. The SPH equations 
derived from a Lagrangian \citep{springel02,monaghan02} yield different 
symmetries in the particle indices, but a recent comparison \citep{rosswog07c} 
between these two sets of equations found only very minor differences. 
The forces from self-gravity are evaluated
via a binary tree \citep{bet1990}, the gravitational forces from the 
central black hole are calculated using a Paczy\'nski-Wiita pseudo 
potential \citep{pa1980}. Details of how the singularity is treated numerically
can be found in \citet{rosswog05a}.
We have taken particular care to avoid artifacts from the use of artificial
viscosity (AV). The so-called Balsara-switch \citep{ba1995} is
implemented here to avoid spurious shear forces. In addition, 
time-dependent viscosity parameters \citep{mo1997} are used so that
AV is only present where really needed. Details of the AV implementation and
tests can be found in \citet{ro2000}.

{\it Equation of state}. The HELMHOLTZ equation of state (EOS),
developed by the Center for Astrophysical Thermonuclear Flashes at the
University of Chicago is used. This EOS accepts an externally 
calculated chemical composition, facilitating the coupling to nuclear 
reaction networks. The electron-/positron equation of state 
makes no assumptions about the degree of degeneracy or relativity
and the exact expressions are integrated numerically and tabulated.
The interpolation in this table enforces thermodynamic consistency by 
construction \citep{timmes00}. The nuclei in the gas are treated as a 
Maxwell-Boltzmann gas, the photons as blackbody radiation.

{\it Nuclear burning}. To account for the feedback onto the fluid from
nuclear transmutations we use a minimal nuclear reaction network
developed by \citet{hi1998}. It couples a conventional $\alpha$-network 
stretching from He to Si with a quasi-equilibrium-reduced $\alpha$-network 
for temperatures in excess of $3 \cdot 10^9$ K. Although a set of only
seven nuclear species is used, this network reproduces all burning
stages from He-burning to nuclear statistical equilibrium (NSE)
accurately. For details and tests we refer to \citet{hi1998}.

\section{Tidal Crushing and Ignition of White Dwarfs}

\begin{figure*}
\plotone{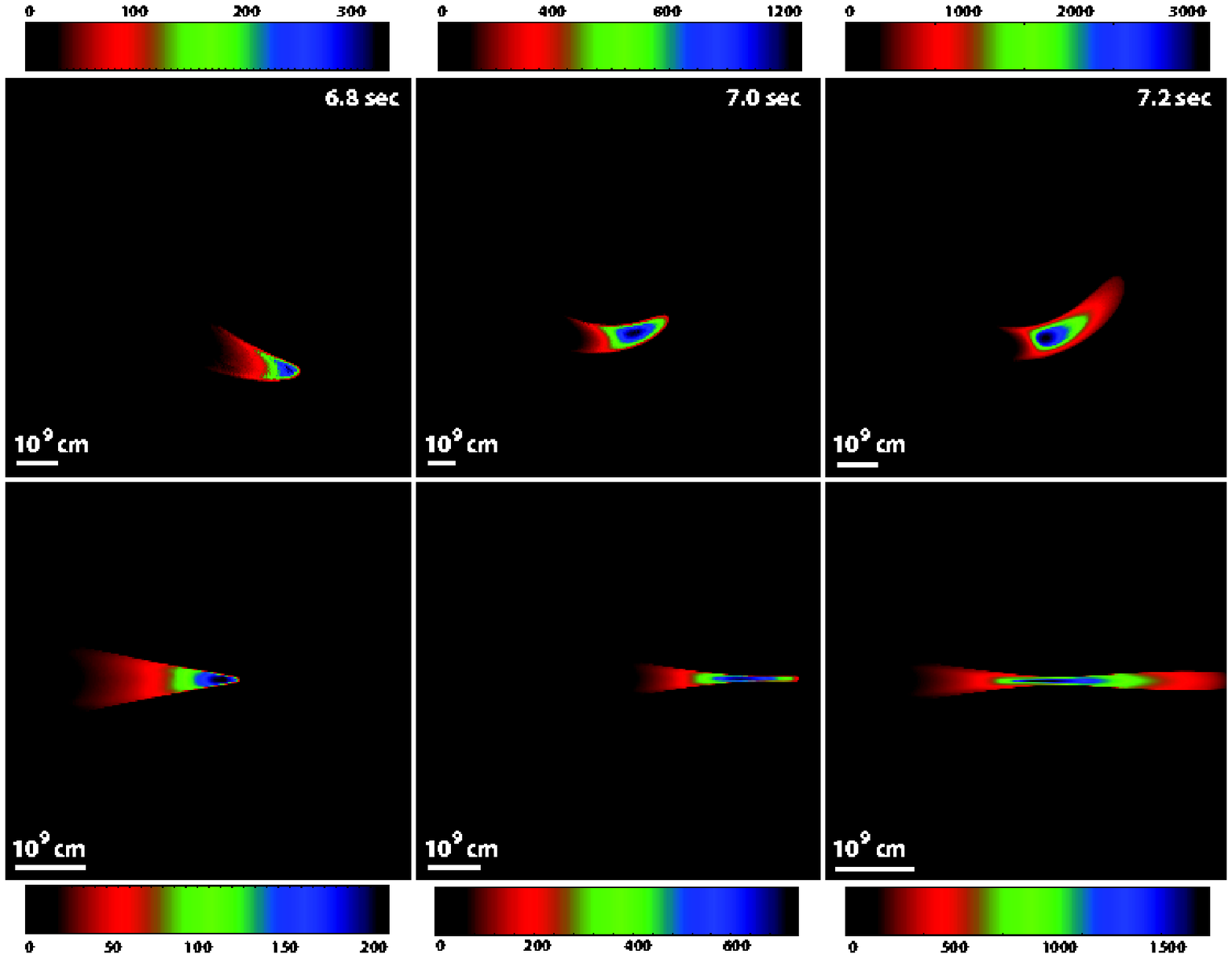}
\caption{A 0.2 $M_\odot$ white dwarf (modeled with more than $4 \times
  10^6$ SPH particles) approaching a $10^3\;M_\odot$ black hole on a
  parabolic orbit with pericenter distance $r_{\rm min} = r_{\rm
    T}/12$ is distorted, spun up during infall and then tidally
  disrupted. Temperatures (in units of $10^6$ K) of the white
  dwarf-black hole encounter are shown.  The panels in the upper row
  show cuts through the orbital (xy-) plane before and after passage
  through pericenter, as the white dwarf attains its maximum degree of
  compression. The panels in the lower row show the temperature
  distribution in the xz-pane (averaged along the y-direction).}
\label{fig1}
\end{figure*}

Several snapshots taken from our numerical simulations of a 0.2
$M_\odot$ white dwarf approaching a $10^{3}M_\odot$ black hole on a
parabolic orbit with pericenter distance well within the tidal radius
($r_{\rm min}= r_{\rm T}/12$) are shown in Figure~\ref{fig1}. While
falling inwards towards the hole, the star develops a quadrupole
distortion which attains an amplitude of order unity by the time of
disruption at $r \sim r_{\rm T}$. The resultant gravitational torque
spins it up to a good fraction of its corotation angular velocity by
the time it gets disrupted. The large surface velocities and the order
unity tidal bulge combine to overcome the star's self gravity and lead
to the disruption of the star.

\begin{figure}
\plotone{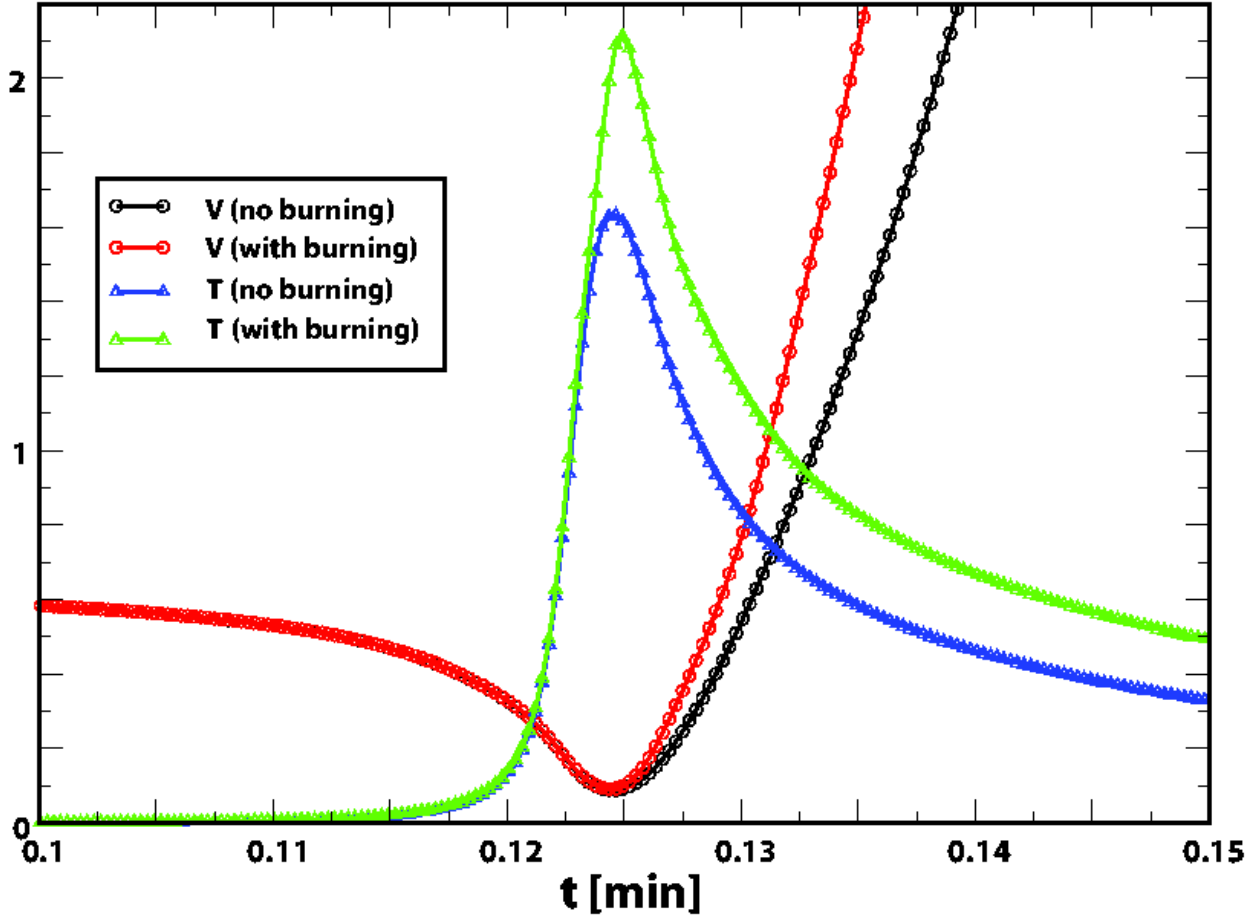}
\caption{Evolution of the thermodynamical and mechanical properties of
  the central portion of the disrupted white dwarf. In this test
  calculation twenty central SPH-particles are chosen and their
  average temperature and volume ($V= \sum_i m_i/\rho_i$ with $m_i$
  particle mass and $\rho_i$ particle density) evolution is followed
  during passage through pericenter. As the star penetrates deeply
  within the tidal radius, the pressure cannot build up sufficiently
  until after the star has undergone an important volume contraction.
  The strong compression of the volume goes along with a sharp
  temperature increase. When burning is taken into account (red and
  green curves), the pressure terms grows very sharply, and, as a
  result, the temperature (averaged over the chosen set of particles)
  increases beyond $2.1\times 10^9$ K.}
\label{fig2}
\end{figure}

The behavior of a white dwarf passing well within the tidal radius
exhibits special features \citep{lu1989}. As illustrated in
Figure~\ref{fig1}, the degenerate star is not only elongated along the
orbital direction but is even more severely compressed into a prolate
shape (i.e., a pancake aligned in the orbital plane). Each section of
the star is squeezed through a point of maximum compression at a fixed
point on the star's orbit. During this very short lived phase, the
pressure grows sharply and the matter is raised to a higher
adiabat. The distortions $\Delta r_{\rm wd}/r_{\rm wd} \sim 1$ impart
supersonic bulk flow velocities during the drastic compression of the
stellar material. As a result, the temperatures increase sharply and
trigger explosive burning (of He for the case shown in
Figure~\ref{fig1}). The corresponding energy release can be very large
(more than the star's self gravity) and it can build up sufficient
pressure to significantly modify the dynamical evolution.
Figure~\ref{fig2} shows the evolution of the compressed, and tidally
disrupted white dwarf with and without the effects of nuclear burning.
The pressure cannot build up sufficiently until after the white dwarf
has endured a substantial volume contraction. This contraction is
halted further by nuclear energy release. Just after the white dwarf
attains its maximum compression, the density drops dramatically as a
result of the subsequent expansion.

During the brief but strong compression phase, the temperature reaches
values beyond $3\times 10^9$ K, approaching but not quite reaching
NSE. As a result, nuclei up to and beyond silicon are synthesized from
the initially pure $^{4}$He 0.2 $M_\odot$ white dwarf. The initial
composition was pure $^{4}$He and the final mass fraction in
iron-group nuclei is roughly 15\%, located near the center
(Figure~\ref{fig3}).  This result should be taken as a modest
underestimate, due to limitations in the seven species nuclear
reaction network.  Post-processing calculations, using a 300 isotope
nuclear network over thermodynamic histories resulting from these
calculations, show significant nuclear flow above silicon, for helium-
rich portions of the gas with peak temperatures above $2
\times10^9$K. As a result, heavier elements (like calcium, titanium
and chromium) would be made, accompanied by a modest increase in the
energy generation.  It is thus safe to conclude that the white dwarf
is tidally ignited and that a sizable mass of iron-group nuclei is
injected into the flow.  Although the thermodynamical evolution and
the nuclear energy release are sensitive to the initial stellar
composition, here assumed to be $^{4}$He, we found that ignition is a
natural outcome for white dwarfs of all masses (e.g. massive white
dwarfs deprived of helium resulting from the evolution of a medium
mass star) passing well within the tidal radius. For example, a C/O
0.6 $ M_\odot$ (1.2 $ M_\odot$) white dwarf approaching a 500 M black
hole on a parabolic orbit with pericenter distance $r_{\rm min} =
r_{\rm T}/5 \; (r_{\rm min} = r_{\rm T}/2.6)$ ignites and, as a
result, at least 0.17 (0.66) $ M_\odot$ of iron-group nuclei are
synthesized in the flow.  Thus, in the most favorable cases, the
nuclear energy release may be comparable to that of typical type Ia
supernovae.

Although the explosion will increase the fraction of ejected debris,
enough remains to be accreted on to the hole (Figure~\ref{fig4}). The
returning gas does not immediately produce a flare of activity from
the black hole. First material must enter quasi circular orbits and
form an accretion torus \citep{re1988,ev1989}. Only then will viscous
effects release enough binding energy to power a flare. The bound
orbits are very eccentric, and the range of orbital periods is
large. For white dwarfs, the orbital semi-major axis of the most
tightly bound debris is $a\sim 300 M_{\rm h,3}^{-1/3} r_{\rm wd,9}
M_{\rm wd,0.6}^{-2/3}\; r_{\rm g}$, and the period is only $t_{\rm
  a}\sim 150 (a/ 300 r_{\rm g})^{3/2} M_{\rm h,3}^{-1/2}$ s.  The
simulation shows that the first material returns at a time $\leq
t_{\rm a}$, with an infall rate roughly given by $10^2 M_\odot\;{\rm
  yr^{-1}}$. Such high infall rates are expected to persist, relative
steadily, for at least a few orbital periods, before all the highly
bound material rains down. The vicinity of the hole would thereafter
be fed solely by injection of the infalling matter at a rate that
drops off roughly as $t^{-5/3}$ for $t \geq t_{\rm fb} \approx 600$
s. Once the torus is formed, it will evolve under the influence of
viscosity, radiative cooling winds and time dependent mass inflow.  A
luminosity $\sim L_{\rm Edd}=10^{41}M_{{\rm h},3}\;{\rm erg\;s^{-1}}$
can therefore only be maintained for at most a year; thereafter the
flare would rapidly fade. It is clear that most of the debris would be
fed to the hole far more rapidly than it could be accepted if the
radiative efficiency were high; much of the bound debris must either
escape in a radiatively-driven outflow or be swallowed
inefficiently. The rise and the peak bolometric luminosity can be
predicted with some confidence. However, the effective surface
temperature (and thus the fraction of luminosity that emerges
predominantly in the soft X-ray band) is harder to predict, as it
depends on the size of the effective photosphere that shrouds the
hole.

\section{Discussion}
Exactly how often a star gets close enough to the central hole to be
tidally disrupted depends on the stellar velocity distribution in star
clusters \citep{fr1976}. In a simple case when the velocities are
isotropic, the frequency with which a star enters the zone of
vulnerability is
\begin{equation}
\sim 10^{-7} M_{\rm h,3}^{4/3}\left({n_* \over 10^{6}\;{\rm
    pc^{-3}}}\right)\left({\sigma \over {\rm 10 \;km\;
    s^{-1}}}\right)^{-1}\left({r_{\rm min} \over r_{\rm
    T}}\right)\;{\rm yr^{-1}},
\label{rate}
\end{equation}
where $n_*$ is the star density in the star cluster nucleus and
$M_{\rm h,3}$ is the hole's mass in units of $10^3 M_\odot$. This
fiducial capture rate may need to be modified for a number of reasons. 
The actual rate could be lower than that given by equation (\ref{rate}),
even if the initial distribution were isotropic, if the {\it loss
  cone} orbits were depleted faster than they could be replenished
\citep{fr1976}. If the relaxation timescale were short enough to
ensure replenishment of the loss cone, then the same diffusion
processes could build up a {\it cusp} in the stellar distribution near
the black hole. Capture or disruption of these stars could then
augment the rate given by equation (\ref{rate}). None of these
complications is, however, likely to seriously modify the simple
estimate given above \citep{ba2004}.
The rate given by equation (\ref{rate}), although highly simplified,
agrees well with the fiducial rates derived from detailed $N-$body
simulations of multimass star clusters containing IMBHs
\citep{ba2004}. However, in these simulations Baumgardt and
collaborators found that IMBHs in star clusters disrupt mainly main
sequence stars and giants, with white dwarfs accounting for only a few
tenths of all disruptions. The probability of a pericentric distance
within $r_{\rm min}$ is proportional to the first power of $r_{\rm
  min}$ (see equation [\ref{rate}]). While complete disruption
requires $r_{\rm min} \leq r_{\rm T}$, ignition, as calculated here,
appears to be a natural outcome for white dwarfs passing well within
the tidal radius: $r_{\rm min} \leq r_{\rm T}/\xi$ with $\xi \geq
3$. The actual rate for white dwarf ignition would then be at least an
order of magnitude lower than that given by equation (\ref{rate}).

\begin{figure*}
\plotone{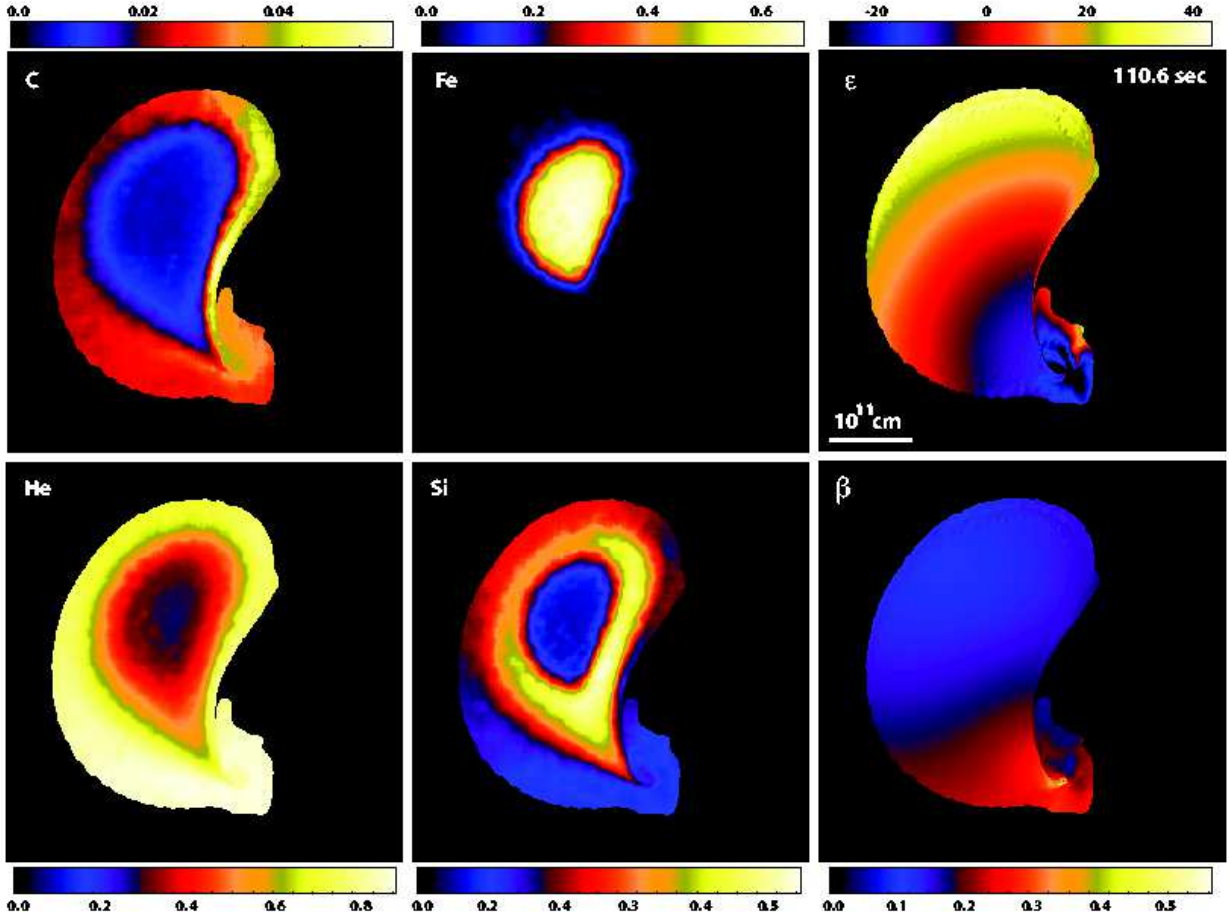}
\caption{The nature of the unbound debris at 110.6s.  The panels show
  cuts through the orbital (xy-) plane well after passage through
  pericenter. The panels in the left columns show the mass fraction
  distributions of the synthesized group elements.  The panels in the
  right column show the velocity distribution, $\beta=v/c$, and the
  specific energy, $\epsilon$, (in code units).}
\label{fig3}
\end{figure*}

\begin{figure}
\plotone{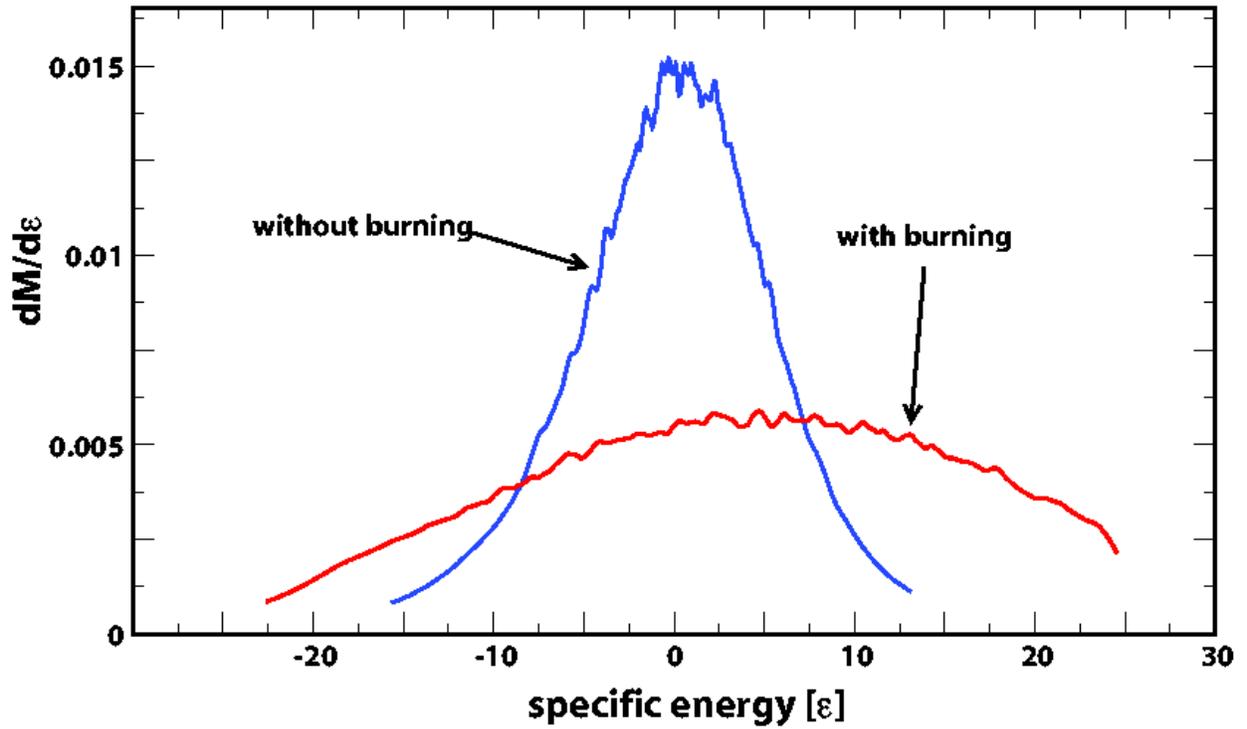}
\caption{Influence of the nuclear burning on the ejected matter:
  differential mass distributions in specific energy for the
  $0.2\;M_\odot$ white dwarf debris. The resultant explosive energy
  severely decreases the amount of material bound to the hole from
  50\% to 35\% of the initial mass of the white dwarf.}
\label{fig4}
\end{figure}

It may seem, however, that even the modest rate of stellar disruptions
given in equation (\ref{rate}) could have conspicuous
consequences. The mass fraction that is ejected rather than swallowed,
though less spectacular than typical Ia supernovae \citep{hi2000},
could nevertheless have distinctive observational signatures. The
ejected material would be concentrated in a cone or {\it fan} close to
the orbital plane (Figure~\ref{fig3}). The characteristic velocities
in excess of $\sim 2 \times 10^4$ km/s (Figure~\ref{fig3}), the
kinetic output when a white dwarf is disrupted being $\sim 4 \times
10^{50}$ erg s$^{-1}$. The light curve powered by radioactive decay,
would indicate the synthesis of at least $0.1 M_{\rm wd}$ of $^{56}$Ni
before the spray of gas becomes translucent. With a rate of disruption
of white dwarfs $\xi \sim 10^{-8}\;{\rm yr^{-1}}$ per globular cluster
\citep{ba2004}, we expect the mean time interval between successive
captures per galaxy to be $\sim 10^{5}(\xi /10^{-8}\;{\rm
yr^{-1}})^{-1} (f_{\rm bh} N_{\rm gc}/ 10^3)^{-1}$ yr, where $f_{\rm
bh} N_{\rm gc}$ is the average number of globular clusters harboring
an IMBH per galaxy and the number of globular clusters per host,
$N_{\rm gc}$, is found, for example, to vary between $5 \times 10^2$
and $7 \times 10^3$ from dwarf ellipticals to giant ellipticals
\citep{va1998,br2006}. Event rates are then estimated to be
significantly less than the rate of type Ia supernovae, which have an
observed time interval between one type Ia explosion and the next of
about $10^2$ yr \citep{le2001,sh2007}, but could be frequent enough to
warrant a search for this new class of optical transient.

The transient sky at faint magnitudes is poorly known, but there are
major efforts under way that would increase the discovery rate for
type Ia supernovae from a few thousands \citep{al2007} to about
hundreds of thousands \citep{ri2006} per year. The transient itself
should have several distinguishing characteristics. First, it should
be seen in association with a globular cluster. Second, the explosion
itself should be different, since the disrupted, degenerate stars
should be, on average, lighter than those exploding as type Ia
supernovae. Third, the spectra should exhibit large Doppler shifts, as
the ejected debris would be expelled with speeds $\ge 10^4$ km/s and
the optical light curve should be rather unique as a result of the
radiating material being highly squeezed into the orbital plane (one
thus expects different timescales for conversion of nuclear energy to
observable luminosity when compared with normal type Ia
events). Finally, the peculiar, underluminous thermonuclear explosion
should be accompanied by a thermal transient (predominantly of soft
X-ray emission) signal with peak intensity $L\sim L_{\rm
  Edd}=10^{41}M_{h,3}$ erg/s, fading within a year.  Such transient
signals, if detected, would be a compelling testimony for the presence of
IMBHs in the centers of globular clusters.

\acknowledgments We thank Holger Baumgardt, Peter Goldreich, Jim Gunn,
Piet Hut, Dan Kasen and Martin Rees for very useful
discussions. The simulations presented in this paper were performed on the JUMP
computer of the H\"ochstleistungsrechenzentrum J\"ulich.
E. R. acknowledges support from the DOE Program for
Scientific Discovery through Advanced Computing (SciDAC;
DE-FC02-01ER41176). W. R. H. has been partly supported by
the National Science Foundation under contracts PHY-0244783 and AST-0653376.
Oak Ridge National Laboratory is managed by UT-Battelle, LLC, for the  
U.S. Department of Energy under contract DE-AC05-00OR22725.

\clearpage

\end{document}